\newcommand*{\ATLASLATEXPATH}{}
\documentclass[cernpreprint,texlive=2016,txfonts,USenglish,texmf]{\ATLASLATEXPATH atlasdoc}
\usepackage[lineno=false]{\ATLASLATEXPATH atlaspackage}
\usepackage{\ATLASLATEXPATH atlasbiblatex}

\usepackage{\ATLASLATEXPATH atlasphysics}

\usepackage{DijetAsymmetry-defs}

\hypersetup{pdftitle={ATLAS draft},pdfauthor={The ATLAS Collaboration}}

\AtlasTitle{Measurement of jet $p_{\mathrm{T}}$ correlations in Pb+Pb and $pp$
collisions at $\sqrt{s_{\mathrm{NN}}}=2.76$~\textrm{Te\kern -0.1em V}
with the ATLAS detector}
\author{The ATLAS Collaboration}

\date{\today}

\PreprintIdNumber{CERN-EP-2017-054}

\AtlasJournalRef{\PLB 774 (2017) 379}
 \AtlasDOI{10.1016/j.physletb.2017.09.078}

\AtlasAbstract{Measurements of dijet \pt\ correlations in Pb+Pb and $pp$
collisions at a nucleon--nucleon centre-of-mass energy of $\sqrt{s_{\mathrm{NN}}}=2.76$~\textrm{Te\kern -0.1em V} are presented. The measurements are performed with the
ATLAS detector at the Large Hadron Collider using
Pb+Pb and \pp\ data samples corresponding to integrated luminosities of 0.14 nb$^{-1}$
and 4.0 pb$^{-1}$, respectively. Jets are reconstructed
using the anti-$k_t$ algorithm with radius parameter values $R=0.3$ and $R=0.4$. A background
subtraction procedure is applied to correct the jets for the large underlying event
present in Pb+Pb collisions. The leading and sub-leading jet transverse momenta are denoted $p_{\mathrm{T_{\mathrm{1}}}}$
and $p_{\mathrm{T_{\mathrm{2}}}}$. An unfolding procedure is applied to the two-dimensional  (\ptone,\,\pttwo) distributions to account for experimental effects in the
measurement of both jets. Distributions of \mbox{$(1/N)\diff N/\diff \xj$}, where
$x_{\mathrm{J}}=p_{\mathrm{T}_{2}}/p_{\mathrm{T}_{1}}$, are
presented as a function of $p_{\mathrm{T_{\mathrm{1}}}}$ and collision centrality. The
distributions are found to be similar in peripheral Pb+Pb collisions
and $pp$ collisions, but highly modified in central Pb+Pb
collisions. Similar features are present in both the $R=0.3$ and
$R=0.4$ results, indicating that the effects of the underlying event are properly accounted for in the measurement. The results are qualitatively consistent with expectations from partonic energy loss models.}

\begin{document}

\maketitle

\tableofcontents

\section{Introduction}
\label{sec:introduction}

Jets have long been considered an important tool for studying the matter
produced in ultra-relativistic heavy-ion collisions. In these
collisions, a hot medium of deconfined colour charges is produced, known
as the quark--gluon plasma (QGP). Jets produced in the initial stage
of the collision lose energy as they propogate through the medium. This
phenomenon, known as \emph{jet quenching}, was first observed at
the Relativistic Heavy Ion Collider (RHIC)~\cite{Adcox:2004mh,Adams:2005dq}. Early measurements using
fully reconstructed jets in \PbPb\ collisions at the LHC provided a
direct observation of this phenomenon~\cite{HION-2010-02}. In \PbPb\
collisions the transverse momentum (\pt) balance between two jets
was found to be distorted, resulting from configurations in which the two jets
suffer different amounts of energy loss. This measurement was the
experimental confirmation of some of the initial pictures of jet
quenching and signatures of a deconfined medium~\cite{Bjorken:1982tu}.

Subsequent measurements of jets in \PbPb\ collisions have improved the
understanding of properties of quenched jets and the empirical
features of the quenching mechanism \cite{HION-2011-02,HION-2013-06,HION-2012-12,
CMS-HIN-10-004,CMS-HIN-11-013,CMS-HIN-11-010,HION-2012-08,CMS-HIN-11-004,CMS-HIN-12-013, Abelev:2013kqa}. Significant
theoretical advances also occurred in this period, and while a
complete description of jet quenching is not available, some models
are capable of reproducing its key features and providing testable
predictions.  Measurements of the dijet asymmetry,
$A_{\mathrm{J}}\equiv
(\ptone-\pttwo)/(\ptone+\pttwo)$,
where \ptone\ and \pttwo\ are the transverse momenta of the
jets with the highest and second highest \pt\ in the event, respectively, have been
crucial in facilitating these developments.  The experimental results
demonstrate that the measured asymmetries in central collisions, where the geometric overlap of the colliding nuclei is almost complete, differ from those in \pp\ collisions more than is expected from detector-specific experimental
effects~\cite{HION-2010-02,CMS-HIN-11-010,CMS-HIN-11-013}. However,
such effects, in particular the resolution of the measured jet \pt,
must be corrected for in order for the measurement to be directly compared to
theoretical calculations. Unfolding procedures have been applied to
correct for such effects for single-jet measurements \cite{HION-2013-06}; however, the dijet
result requires a two-dimensional unfolding to account for migration
in the \pt\ of each jet separately. The measurement reported here
is the first unfolded \PbPb\ dijet measurement and as such can be directly compared to theoretical models.

This \papertype\ presents a measurement of dijet \pt\ correlations in
\PbPb\ and \pp\ collisions at a nucleon--nucleon centre-of-mass energy
of 2.76~\TeV\ performed with the ATLAS
detector.  Jets are reconstructed with the anti-$k_t$ algorithm with radius parameter values $R=0.3$ and $R=0.4$ \cite{Cacciari:2008gp}. The analysis is described mostly for the example of $R=0.4$ jets. A background subtraction procedure is applied to account for
the effects of the large underlying event (UE) present in \PbPb\
collisions on the measured jet kinematics. The momentum balance of the dijet system is expressed by the
variable $\xJ\equiv \pttwo/\ptone$. Measurements of the dijet yield normalised by the total number of jet pairs in a given \ptone\ interval, \dNdxJ, are
presented as a function of \xJ\ in intervals of \ptone\ and collision centrality. The
results are obtained by first measuring the two-dimensional distribution, \ptonepttwo,
and unfolding in the two-dimensional space. The binning in the
\ptonepttwo\ distribution is chosen such that the bins in the
two-dimensional space correspond to fixed ranges of \xJ, and the \dNdxJ\ results
are obtained by projecting into these \xJ\ bins. The \ptonepttwo\ distributions are less strongly correlated for jets reconstructed with a
smaller value of $R$ due to the effects
of parton radiation outside the jet cone, which makes them less suitable as a probe of medium-induced
effects in \PbPb\ collisions. However, for smaller jet sizes the effect of
the UE on the measurement is significantly reduced. It is
therefore interesting to compare the results obtained using $R=0.3$
and $R=0.4$ jets, to see if the same features are visible.

\section{Experimental set-up}
\label{sec:detector}
The measurements presented in this \papertype\ are
performed using the ATLAS inner detector, calorimeter and trigger systems
\cite{PERF-2007-01}. The inner detector provides
measurements of charged-particle tracks over the range $\Aeta<2.5$.\footnote{\atlasfootnote} It is composed of silicon pixel detectors
in the innermost layers, followed by silicon microstrip detectors and
a straw-tube tracker, all immersed in a 2~T axial magnetic field
provided by a solenoid. The minimum-bias trigger
scintillators (MBTS) measure charged particles over $2.1 < \Aeta <
3.9$ using two planes of counters placed
at $z = \pm 3.6$~m and provide timing measurements used in
the event selection~\cite{HION-2011-05}.

The ATLAS calorimeter system consists of a liquid argon (LAr)
electromagnetic (EM) calorimeter ($\Aeta<3.2$), a steel--scintillator
sampling hadronic calorimeter ($\Aeta < 1.7$), a LAr hadronic
calorimeter ($1.5 < \Aeta < 3.2$), and a forward calorimeter (FCal)
($3.2 < \Aeta < 4.9) $.  The hadronic calorimeter has three sampling layers longitudinal in shower depth and has a $\Delta \eta \times \Delta
\phi$ granularity of $0.1 \times 0.1$ for $\Aeta < 2.5$ and $0.2
\times 0.2$ for $2.5 < \Aeta < 4.9$.\footnote{An exception is the third
sampling layer, which has a segmentation of $0.2 \times 0.1$ up to
$\Aeta = 1.7$}  The EM calorimeters are longitudinally segmented in shower depth 
into three compartments following a pre-sampler layer ($\Aeta<1.8$). The EM
calorimeter has a granularity that varies with layer and
pseudorapidity, but which is generally much finer than that of the
hadronic calorimeter. The first layer has high $\eta$ granularity
(between 0.003 and 0.006) that can be used to identify photons and
electrons. The middle sampling layer, which typically has the largest
energy deposit in EM showers, has a granularity of $0.025 \times
0.025$ over $\Aeta < 2.5$. A total transverse energy (TE) trigger is
implemented by requiring a hardware-based determination of the total transverse energy in the
calorimeter system, \TE, to be above a threshold. 

The zero-degree calorimeters (ZDCs) are located symmetrically at $z =
\pm 140$~m and cover $\Aeta > 8.3$. In \PbPb\ collisions the ZDCs
primarily measure ``spectator'' neutrons: neutrons that do not interact hadronically 
when the incident nuclei collide. A ZDC coincidence trigger is
implemented by requiring the pulse height from each ZDC to be
above a threshold set below the single-neutron peak. 

In addition to the ZDC and TE hardware-based triggers, a
software-based high-level trigger is used to further reduce
the accepted event rate. This trigger applies a jet reconstruction
procedure, including a UE subtraction, similar to that used in the
offline analysis, which is described in Section 4.

\section{Data and Monte Carlo samples}
\label{sec:samples}

The \PbPb\ data used for these measurements were recorded in 2011 and obtained using a
combination of jet and minimum-bias triggers. The minimum-bias
trigger is defined by a logical OR of the TE trigger with a threshold
of $\TE = 50$~\GeV\ and the ZDC coincidence trigger. The combined trigger is fully efficient in the range of centralities presented
here. In the events selected by the ZDC coincidence trigger alone, at least one track is required to remove empty events.
The jet trigger~\cite{ATLAS:2016qun} first selects events satisfying the TE trigger with a
threshold of $\TE = 20$~\GeV. A jet reconstruction procedure is then applied using the anti-\kt\
algorithm with $R=0.2$ and utilising a UE subtraction procedure
similar to that used in the offline reconstruction described in
Section~\ref{sec:reconstruction}. Events with at least one
jet with $\ET > 20$~\GeV\ at the electromagnetic
scale~\cite{PERF-2011-03} are selected by the jet trigger. The use of $R=0.2$ for jets in the trigger,
as opposed to the values of $R=0.3$ and $0.4$ applied in the measurement, is
motivated by the need to define an algorithm that is robust against
UE fluctuations, which grow with $R$. The effects of the different $R$ values on the trigger efficiency are discussed in
Section~\ref{sec:analysis}. The minimum-bias trigger operated with a prescale of approximately 18
while no prescale was applied to the jet trigger. After accounting for
these prescales, the recorded
events correspond to integrated luminosities of 8\invmub\ and
0.14\invnb\ for the minimum-bias and jet-triggered samples,
respectively.

Events are further subjected to criteria designed to remove
non-collision background and inelastic electromagnetic interactions between the nuclei. Events are
required to have a reconstructed primary vertex and have a timing
difference of less than 5 ns between the times measured by the two
MBTS planes. After the trigger and event selection criteria, the
resulting data samples contain 53 and 14 million events in the
minimum-bias and jet triggered samples, respectively. The average number of collisions
per bunch-crossing in the \PbPb\ data sample was less than
0.001, and the effects of multiple collisions are neglected in the
data analysis.

The centrality of the \PbPb\ collisions is characterised by the total
transverse energy measured in the FCal modules, \ETfcal. The \ETfcal\
distribution obtained in minimum-bias collisions is partitioned into
separate ranges of \ETfcal\ referred to as centrality
classes~\cite{Miller:2007ri,Alver:2008aq,HION-2011-05}. Each class is
defined by the fraction of the distribution contained by the interval,
e.g. the 0--10\% centrality class, which corresponds to the most central collisions, contains the 10\% of minimum-bias
events with the largest \ETfcal. The centrality boundaries used in
this analysis are 0\%, 10\%, 20\%, 30\%,  40\%, 60\% and 80\%.

The \pp\ data sample, recorded in 2013, was composed of events
selected by a jet trigger and used a series of different \pt\
thresholds each selected with a different prescale. The jet trigger is
the same used in other ATLAS measurements in \pp\ collisions~\cite{ATLAS:2016qun} and applies the anti-\kt\ algorithm
with $R=0.4$. The events are further
required to contain at least one primary reconstructed vertex. The average number of \pp\ collisions
per bunch-crossing varied between 0.3 and 0.6 during data
taking. The sample corresponds to a luminosity of 4.0\invpb.

The impact of experimental
effects on the measurement is evaluated using the
\textsc{Geant4}-simulated detector response~\cite{Agostinelli:2002hh,SOFT-2010-01} in a Monte Carlo (MC) sample of
\pp\ hard-scattering events.  Dijet events at $\sqrts= 2.76$~\TeV\
are generated using \textsc{Pythia} version 6.423
\cite{Sjostrand:2006za} with parameter values chosen according to the
AUET2B tune~\cite{mctunes} using the CTEQ6L1 parton distribution function (PDF) set~\cite{Pumplin:2002vw}. To fully populate the kinematic range
considered in the measurement, hard-scattering events are generated
for separate intervals of $\hat{\pt}$, the transverse momentum of
outgoing partons in the $2\rightarrow2$ hard-scattering, and combined
using weights proportional to their respective cross
sections. Separate samples are generated
for the \PbPb\ and \pp\ analyses, with the simulated detector
conditions chosen to match those present during the recording of the
respective data samples.  In the \pp\ data sample, the contribution of
additional collisions in the same bunch crossing (pile-up) is
accounted for by overlaying minimum-bias \pp\ collisions produced at the same rate as in the data, generated by Pythia
version 8.160 \cite{Sjostrand:2007gs} using the A2 \cite{ATL-PHYS-PUB-2012-003} tune
with CT10 PDF set \cite{Lai:2010vv}. In the \PbPb\ sample, the UE contribution to
the detector signal is accounted for by overlaying the simulated events
with minimum-bias \PbPb\ data. The vertex position of each simulated
event is selected to match the data event that is overlaid. Through this procedure the MC sample contains contributions
from underlying-event fluctuations and harmonic flow that match those
present in the data. The
combined signal is then reconstructed using the same procedure as is
applied to the data. So-called \emph{truth jets} are defined by
applying the anti-\kt\ algorithm with $R=0.3$ and $R=0.4$ to stable particles in the MC event generator's
output, defined as those with a proper lifetime greater than 10 ps, but excluding muons and neutrinos, which do not leave significant energy
deposits in the calorimeter. 

The detector's response to quenched jets is studied with an
additional sample using \textsc{Pyquen}~\cite{Lokhtin:2005px}. This event generator applies medium-induced energy loss to parton
showers produced by \textsc{Pythia}. It is used to generate a sample
of jets with fragmentation functions that differ from those in the
nominal \textsc{Pythia} sample in a fashion consistent with
measurements of fragmentation functions in quenched
jets~\cite{HION-2012-08,CMS-HIN-11-004,CMS-HIN-12-013}.

\section{Jet reconstruction}
\label{sec:reconstruction}

The procedure used to reconstruct jets in heavy-ion
collisions is described in detail in Ref.~\cite{HION-2011-02} and is
briefly summarised here. First, energy deposits in the calorimeter
cells are assembled into $\Delta\eta \times
\Delta\phi=0.1\times\frac{\pi}{32}$ logical towers. Jets are formed
from the towers by applying the anti-$k_t$ algorithm
\cite{Cacciari:2008gp} as implemented in the \verb=FastJet= software
package \cite{Cacciari:2011ma}.

An estimate of the UE contribution to each tower within the jet is
performed on an event-by-event basis by estimating the 
transverse energy density, $\rho(\eta,\phi)$. Global azimuthal modulation in
the UE arises due to the physics of flow and is traditionally
described in terms of the Fourier expansion of the $\phi$ dependence of
the transverse energy density. In the subtraction procedure, the UE
estimate is assigned a $\phi$ dependence using the measured magnitudes and
phases of the modulation:
\begin{equation}
\rho(\eta,\phi) =\overline{\rho}(\eta)\times\left(1+2\sum_{n} v_n \cos
[n(\phi-\Psi_n)]\right)\,,
\label{eqn:UE_subtr}
\end{equation}
where $v_n$ and $\Psi_n$ are the magnitudes and phases of the harmonic
modulation, respectively, and $\overline{\rho} (\eta)$ is the average
transverse energy density measured from energy deposits in the
calorimeter as a function of $\eta$.  In Ref.~\cite{HION-2011-02},
only the second-order harmonic modulation ($n=2$) was considered, but
in this measurement the procedure has been extended to account for
$n=3$ and 4 harmonic modulations as well. The subtraction is applied
to each tower within the jet. The quantities in Eq.~(\ref{eqn:UE_subtr})
may be biased if the energy in a jet is included in their
calculation, which results in an over-subtraction of the average UE
contribution to the jet energy or incomplete removal of the harmonic
modulation. To mitigate such effects, the contribution from jets
is excluded from the estimate of the background. The typical background energy subtracted
from the jets varies from a few \GeV\ in peripheral collisions to
150~\GeV\ in the most central collisions.

A calibration factor, derived from MC studies, is then
applied after the subtraction to account for the non-compensating hadronic response. 
A final \emph{in situ} calibration is applied to
account for known differences in detector response between data and the MC sample used to derive the initial
calibration~\cite{PERF-2012-01}. This calibration is derived in 8~\TeV\
\pp\ data and adapted to the different beam energy and pile-up
conditions relevant for the samples considered here. It uses the
balance between jet pairs in different $\eta$ regions of the
detector to provide an evaluation of the relative response
to jets as a function of $\eta$. It subsequently uses jets
recoiling against objects with an independently-determined energy scale such as $Z$ bosons or
photons to provide constraints on the absolute energy measurement.

\section{Data analysis}
\label{sec:analysis}
In this analysis, jet pairs are formed from the
two highest-\pt\ jets in the event with $\pt >25$~\GeV\ and $ \Aeta <
2.1$. The pair is required to have $\Delta \phi > 7\pi / 8$, where $\Delta \phi \equiv |\phi_1 - \phi_2|$. For
events selected by a jet trigger, the leading jet is required to match a jet
identified by the trigger algorithm responsible for selecting the
jet. The two-dimensional \ptonepttwo\ distributions obtained from
different triggered samples are combined such that intervals of
\ptone\ are populated by a single trigger. In the \pp\ data analysis,
the trigger with the most events that is more than 99\% efficient for
selecting a jet with $\pt > \ptone$ is used, with the reciprocal of
the luminosity for the respective trigger samples used as a weight.

The \PbPb\ jet trigger efficiency has a broad turn-on as a function
of \pt\ since the trigger jets are identified using
$R=0.2$ and have no energy scale calibration applied. This effect is
the strongest in central collisions where the UE fluctuations are the
largest and further weaken the correlation between jets reconstructed
with different values of $R$. In the most central collisions, the single-jet-trigger efficiency does not reach a plateau until $\pt \sim 90$~\GeV. The jet-triggered 
sample is used where the efficiency is found to be greater than 97$\%$, which 
occurs at a \pt\ of approximately 85~\GeV\ in the most central collisions. A trigger efficiency 
correction is applied in the region where there is an inefficiency.

In addition to the dijet signal, the measured \ptonepttwo\ distribution
receives contributions from so-called \emph{combinatoric} jet pairs. Such
pairs arise when two jets,  which are not from the same hard-scattering process, fulfill the pair
requirements through random association. Jets forming such pairs may originate from
independent hard scatterings or from upward UE fluctuations identified as jets, referred to as UE jets. The rate for such occurrences is highest in the most central collisions, and the reduction in the true sub-leading jet \pt\
due to quenching effects further enhances the likelihood of forming a
combinatoric pair.

The shape of the $\Delta \phi$ distribution for the combinatoric jet pairs
is influenced by the harmonic flow. Since the jet \pt\ spectrum falls steeply, the jets most likely to be
measured at a given \pt\ value are those lying on top of larger-than-average UE. If the effects of the modulation of the UE are not fully accounted for in the background subtraction, more
jets would be observed at angles corresponding to the flow
maxima ($\phi\sim \Psi_n$). Thus combinatoric jet pairs, without any underlying angular
correlation, are expected to acquire a modulation to their $\Delta \phi$ distribution
determined by the dominant flow harmonics~\cite{HION-2011-01}. Although the second-, third-
and fourth-order harmonic modulations are considered event-by-event in the jet reconstruction procedure described in
Section~\ref{sec:reconstruction}, only the effects of the second-order
modulation on the $\Delta \phi$ distribution are observed to be completely
removed. The residual effects are an indication that the
method of estimating the modulation of the UE underneath the jet is
less accurate for the higher-order harmonics than for $n=2$. 

To account for the residual modulation, the combinatoric contribution
is assumed to be of the form:
\begin{equation}
C(\Delta\phi)=Y (1 + 2
c_{3}\cos3\Delta\phi + 2c_{4}\cos4\Delta\phi)\,.
\label{eqn:combinatoric}
\end{equation}
 The $c_{3}$ and $c_{4}$ values are determined by fitting the $\Delta \phi$
distributions over the range $0 < \Delta \phi < \pi/2$ where the real
dijet contribution is expected to be small. The region $0 < \Delta
\phi \lesssim 0.8$ is also expected to receive real dijet contributions
arising from parton radiation which results in pairs of jets at nearby angles. To remove this contribution, the fit to obtain $c_{3}$ and
$c_{4}$ is performed only using jet pairs with a separation of \etasep. Once $c_3$ and $c_4$ are obtained, the $\Delta \phi$
distribution without this $\Deta$ requirement is integrated over the
range $1 < \Delta \phi < 1.4$ to obtain $Y$. This procedure is performed
separately in each \ptonepttwo\ interval. In intervals where the $c_3$ and $c_4$ are found to not be statistically
significant their values are taken to be zero. The expected combinatorial contribution, $B$,
in the signal region is obtained by integrating $C(\Delta\phi) $ from
$ 7 \pi/8$ to $\pi$. 

\begin{figure}
\centering
\includegraphics[width=0.5\textwidth]{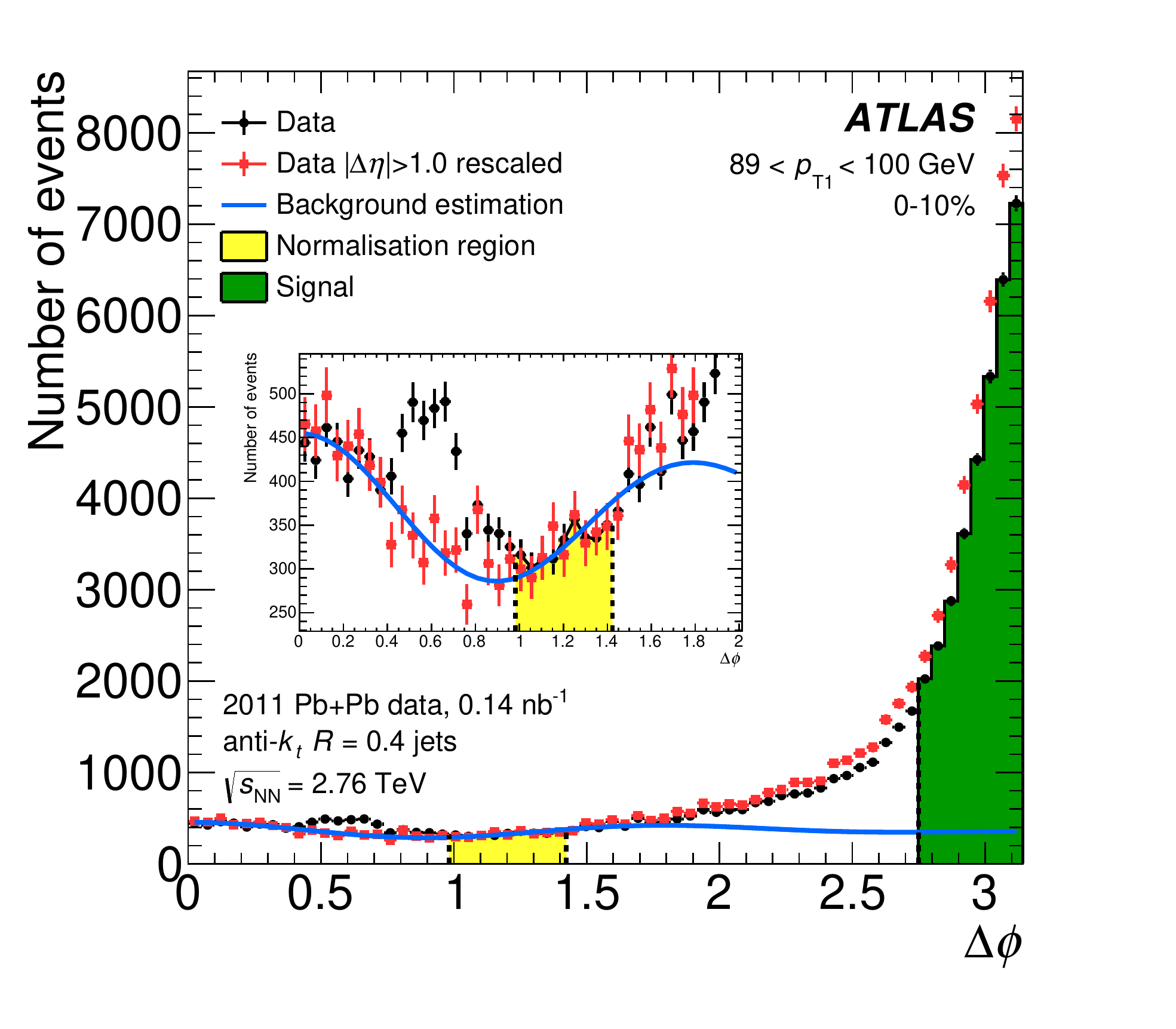}
\caption{The $\Delta \phi$ distribution for $R=0.4$ jet pairs with 89 < \ptone < 100~\GeV\ in the
0--10\% centrality interval. The distribution for all jet pairs is
indicated by the black circles. The combinatoric contribution given by
Eq.~(\ref{eqn:combinatoric}) is shown as a blue line.  The ranges of $\Delta \phi$ used to fix the value of
$Y$ and to define the signal region ($\Delta \phi > \frac{7\pi}{8}$) are
indicated by yellow and green shaded regions, respectively. The parameters $c_3$ and $c_4$ are obtained by fitting the $\Delta \phi$
distribution for jet pairs with \etasep\ in the region $0 <
\Delta \phi < \frac{\pi}{2}$, which is indicated by the red squares (scaled to
match the black circles in the yellow region for presentation purposes). The
error bars denote statistical errors.}
\label{fig:dphi_subtraction}
\end{figure}

The $\Delta \phi$  distribution of jet pairs is
shown in Figure~\ref{fig:dphi_subtraction} for pairs with 89 < \ptone < 100~\GeV\ in the
0--10\% centrality interval. Also shown is the $\Delta \phi$ distribution
obtained from such jet pairs with \etasep, which is fitted to obtain
$c_3$ and $c_4$. 
The
background subtraction is most significant in central collisions, where
the fraction subtracted from the total yield in the signal region is
as large as 10$\%$ for small \xJ\ and is less than 1$\%$ for \xJ\
values greater than 0.5. The background contribution in more
peripheral collisions is less than 1$\%$  for all values of \xJ. This background subtraction is not applied in the \pp\ data because the pile-up is small. 

The presence of combinatoric jet pairs also reduces the efficiency for genuine pairs. The measured inclusive jet spectrum is used to estimate the likelihood that
another jet in the event, uncorrelated with the dijet system, is
measured with a transverse momentum greater than $\pttwo$. For the 40--60\% and 60--80\% centrality intervals
the effect is negligible. In the 0--10\% centrality bin the efficiency is
approximately 0.9 for $\pttwo=25$~\GeV\ and increases with $\pttwo$,
reaching unity at 45~\GeV. The effects of the combinatoric jet pairs are
accounted for by first subtracting the estimated background and then
correcting for the efficiency, $\varepsilon$, in each \ptonepttwo\ bin. The number of jet pairs corrected for such
effects is defined to be:
\begin{equation*}
\Ncorr=\frac{1}{\varepsilon}\left(\NcorrTL - B\right),
\end{equation*}
where \NcorrTL\ is the number of jet pairs after correcting for trigger
efficiency and luminosity/prescale weighting as described above.

\begin{figure}
\centering
\includegraphics[width=0.95\textwidth]{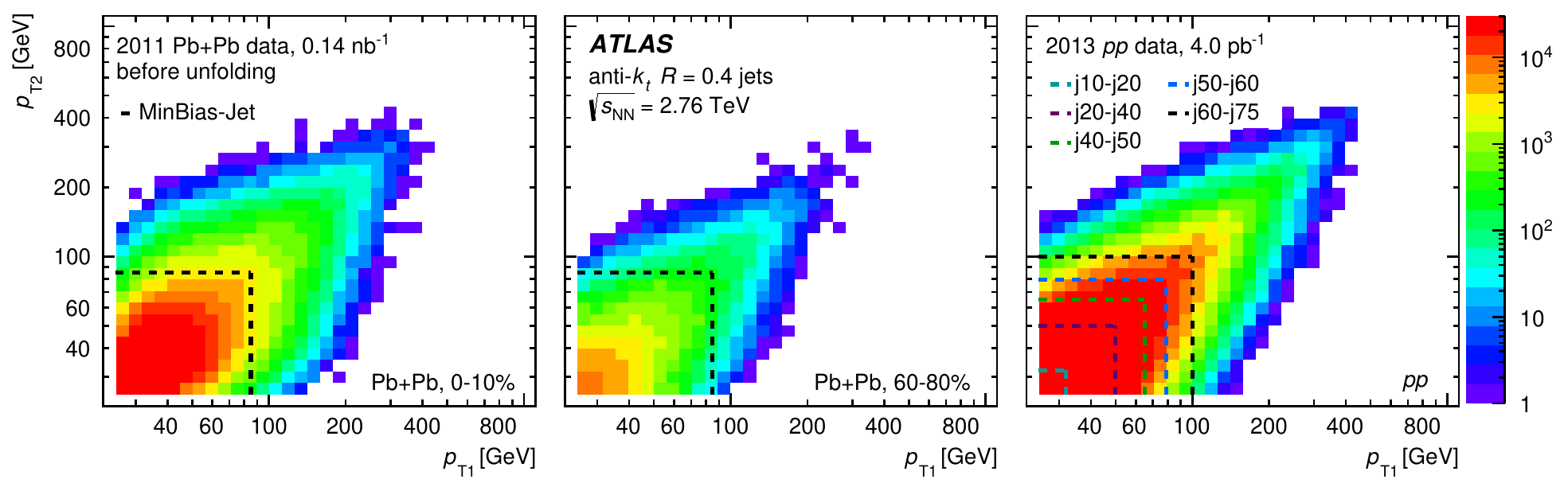}
\caption{The two-dimensional \ptonepttwo\ distributions after correction and symmetrisation for
  \PbPb\ data in the 0--10\% (left) and 60--80\% (centre) centrality
  bins and for \pp\ data (right) for $R=0.4$ jets. The dashed lines indicate the
  boundaries used in selecting the different triggers. The \PbPb\ data
  distributions have their combinatoric contribution subtracted.}
\label{fig:raw_2D_data}
\end{figure}

In a given event, the \pt\ resolution may result in
the jet with the highest true \pt\ being measured with
the second highest \pt\ and vice-versa. To properly account for such
migration effects, \ptonepttwo\ distributions are symmetrised prior
to the unfolding by apportioning half of the yield in a given
\ptonepttwo\ bin, after combinatoric subtraction, to the bin related to the original by $\ptone
\leftrightarrow \pttwo$. The two-dimensional distributions after symmetrisation
are shown in Figure~\ref{fig:raw_2D_data}
for central and peripheral \PbPb\ collisions and for \pp\
collisions. The choice of binning in \ptonepttwo\ is motivated by the
mapping to the \xJ\ variable, and is described in more detail in the following section.

\section{Unfolding}
\label{sec:unfolding}

The calorimetric response to jets is evaluated in
the MC sample by matching truth and reconstructed jets; the nearest
reconstructed and truth jets within $\Delta R=\sqrt{(\Delta \eta)^2 +
  (\Delta \phi)^2}$ of $0.3$ are considered to be a match. The same
requirement is applied in both the $R=0.3$ and $R=0.4$ versions of the analysis. The response
is typically characterised in terms of the jet energy scale (JES) and
jet energy resolution (JER). These quantities describe the mean and
width of the \pTreco\ distributions at fixed
\pTtruth, expressed as a fraction of \pTtruth. Generally, the mean of \pTreco\ differs from \pTtruth\ by less than a
percent, independent of \pTtruth\ and centrality. This
indicates that the subtraction of the average UE contribution to the
jet energy is under good experimental control. The JER receives
contributions both from the response of the calorimeter and from local
UE fluctuations about the mean in the region of the jet. The latter
contribution dominates at low \pt\ with the resolution as
large as 40\% at $\pt\simeq 30$~\GeV\ in the most central
collisions. At the same \pt,  the JER is only 20\% in peripheral collisions,
similar to that in $\textit{pp}$ collisions. At larger \pt\ values the relative contribution of the UE fluctuations to the jet \pt\ diminishes, and the JER is dominated by
detector effects, reaching a constant, centrality-independent value of
8\% for $\pt > 300$~\GeV.

The migration in the two-dimensional \ptonepttwo\
distribution is accounted for by applying a two-dimensional Bayesian
unfolding to the data~\cite{D'Agostini:1994zf, Adye:2011gm}. This procedure utilizes a response matrix
obtained by applying the same pair selections to the truth jets in MC simulation as in
the data analysis (except the trigger requirement) and recording the
values of \ptoneTruth\ and \pttwoTruth\
and the transverse momenta of the corresponding reconstructed jets
\ptoneReco\ and \pttwoReco. The matched
reconstructed jets are not required to have the highest \pt\ in the
event, but are subject to all other requirements applied to the data
and truth jets. The response matrix is populated symmetrically in both truth and reconstructed \pt. The
full four-dimensional response behaves similarly to the factorised
product of separate single-jet response distributions, and the
migration effects can be understood in terms of the above
discussion. While this provides intuition for the nature of the
unfolding problem, such a factorisation is not explicitly assumed, and any correlations between the response of the two jets are accounted for in the procedure.

After unfolding, the leading/sub-leading distinction is restored by
reflecting the distribution over the line $\ptone=\pttwo$: for each
bin with $\pttwo > \ptone$ the yield is moved to the corresponding
bin with $\pttwo < \ptone$. The bins along the diagonal, e.g. those containing pairs with $\pttwo=\ptone$, are not
affected by this procedure. The two-dimensional distribution is constructed
using binning along each axis such that the upper edge of the
$i^{\mathrm{th}}$ bin obeys,
\begin{equation*} p_{\mathrm{T}\,i} = p_{\mathrm{T}\,0}\,
\alpha^{i}\,,\quad \alpha=\left(\frac{ p_{\mathrm{T}\,N}
}{p_{\mathrm{T}\,0}}\right)^{1/N}\,,
\end{equation*} where $N$ is the total number of bins and
$p_{\mathrm{T}\,0}$ and $p_{\mathrm{T}\,N}$ are the minimum and
maximum bin edges covered by the binning, respectively.  As a
consequence, the bins are of the same size when plotted with logarithmic
axes.  With these choices of binning, the range of \xJ\ values in any given
\ptonepttwo\ bin is fully contained within two adjacent \xJ\
bins, which have boundaries at $x_{\mathrm{J}\, i}=\alpha^{i-N}$. In this analysis, half of the yield in each \ptonepttwo\ bin
is apportioned to each of the \xJ\ bins. The exceptions are the bins
along the diagonal. These bins contribute solely to the \xJ\ bin with
bin edges $(\alpha^{-1}, 1)$. The effects of such a mapping on the
\xj\ distribution are studied and found to not significantly distort
the shape of the distribution for a variety of input \xJ\
distributions. 

The Bayesian unfolding method is an iterative procedure that requires
both a choice in a number of iterations, \niter, and
assumption of a prior for the underlying true
distribution. An increase in \niter\ reduces
sensitivity to the choice of prior but may amplify statistical fluctuations that are already present
in the input distribution. As \textsc{Pythia} does not include the effects of jet
quenching, the \xJ\ distributions obtained from the MC sample are not
expected to be optimal choices for the prior. In particular, the \xJ\
distributions in \textsc{Pythia} increase monotonically with \xJ,
whereas the distributions in the data become flatter and develop a peak
near $\xJ\sim0.5$ in lower \ptone\ intervals and in the most central
collisions. The \ptonepttwo\ distributions from \textsc{Pythia} are reweighted in a centrality-dependent way to
obtain features that qualitatively match those present in the
data.

\begin{figure}
\centering
\includegraphics[width=0.95\textwidth]{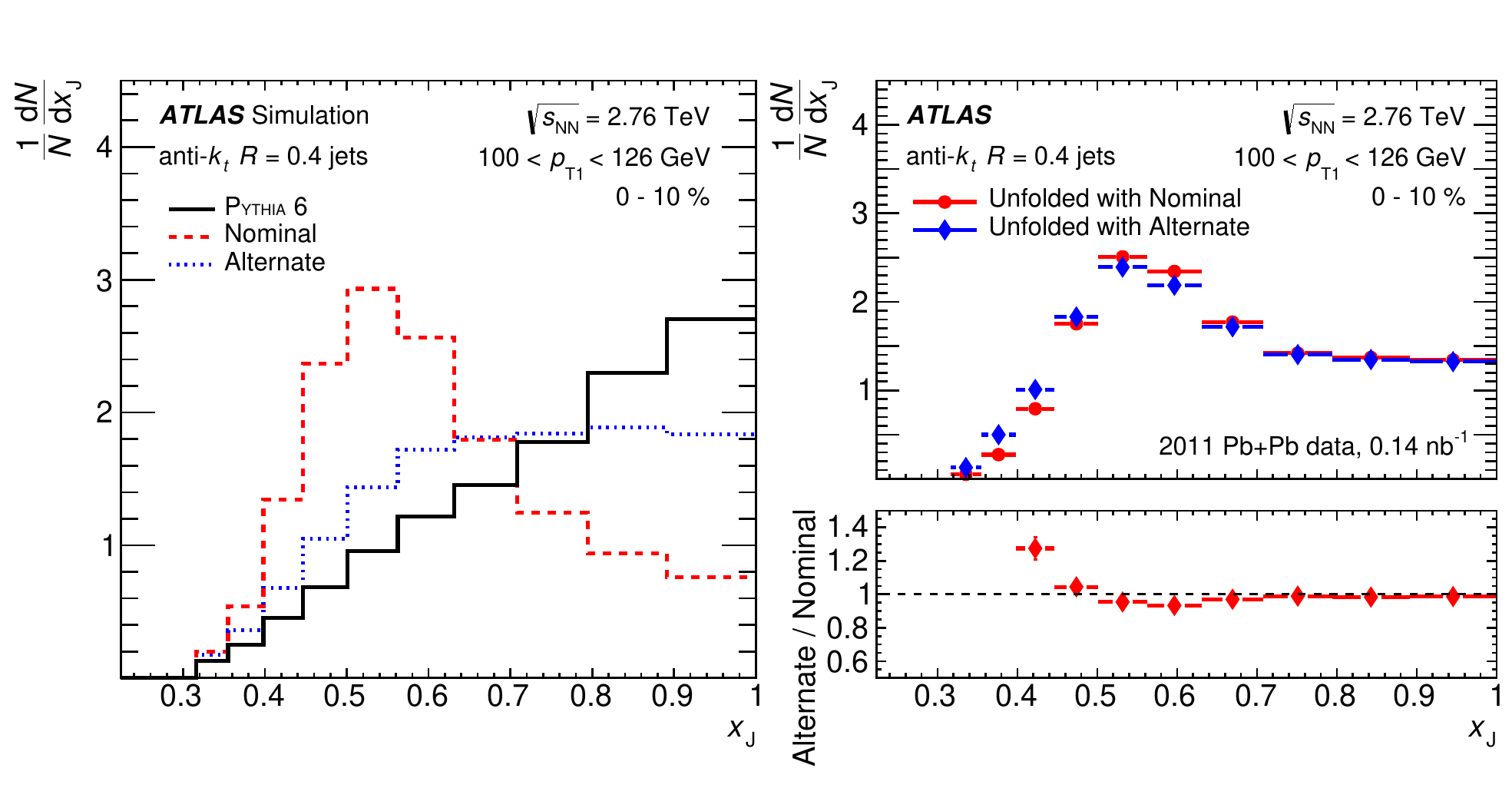}
\caption{Left: the \dNdxJ\ distributions used as priors in the
  unfolding of the $R=0.4$ jets for the nominal (dashed red) and alternate variation (dotted blue) for the $100 < \ptone < 126$~\GeV\ and
0--10\% centrality interval. The same distribution obtained from the
\textsc{Pythia} MC sample is shown in solid black. Right: unfolded \dNdxJ\
distributions from data for the same
\ptone\ and centrality ranges using the nominal (red circles) and
alternate (blue diamonds) priors shown in the left panel. The ratio of nominal to alternate is shown in the bottom panel. In the bottom panel on the right the first two bins are off scale with bins centers of \xJ=0.34 and 0.38 and bins contents of 2.49 and 1.82, respectively. Statistical
errors are not shown.}
\label{fig:priors}
\end{figure}

The effects of the reweighting procedure are shown in the
left panel of Figure~\ref{fig:priors} in the $100
< \ptone < 126$~\GeV\ range and 0--10\% centrality interval, where the
largest difference between the data and \textsc{Pythia} is
observed. The ``nominal'' distribution, or the reweighted distribution, is used as the prior in the
unfolding of the data. An ``alternate'' reweighting is
also shown, which has a shape significantly different from the
nominal, but does not increase as much as the
\textsc{Pythia} distribution. The features in the data
are observed to be robust with respect to the choice of prior for a
broad set of reweighting functions.  The systematic uncertainty due to the choice
of prior is estimated by comparing the results of the unfoldings using
the ``nominal" and ``alternate'' \xJ\ distributions. The results of applying unfoldings with these two choices of priors are shown in the right
panels of Figure~\ref{fig:priors} for the same \ptone\ and centrality selection.

An alternative study is performed in the MC sample to validate the
estimation of this uncertainty. The ``alternate" reweighting is
applied to obtain input truth and reconstructed distributions in which no peak
structure is present. The reconstructed distribution is then
unfolded using the \emph{nominal} prior. The unfolded distribution does not develop the strong peak
present in the nominal prior. The differences between the unfolded result
and the input truth distribution are similar to the uncertainty obtained by varying the prior used to
unfold the data.

The value of \niter\ is selected separately in each centrality
interval by examining the
uncertainty, $\sqrt{\Sigma \delta^2}$, in \dNdxJ\ after unfolding considering statistical uncertainties and
systematic uncertainties attributed to the unfolding procedure,
\begin{equation*}
\delta^2 = \delta^2_{\mathrm{stat}} + \delta^2_{\mathrm{prior}} \,,
\end{equation*}
and summing over all \xJ\ bins. Here $ \delta_{\mathrm{prior}}$ is the uncertainty due to the choice of
prior, obtained using the procedure described above. The statistical uncertainties are
evaluated using a pseudo-experiment technique. Stochastic variations
of the data are generated based on its statistical uncertainty and
each variation is unfolded and projected into \xJ. The statistical covariance of the set
is taken as the statistical uncertainty. An additional covariance is
obtained from applying the pseudo-experiment procedure to the response
matrix and combined with that obtained from applying the procedure to
the data. The $\delta^2_{\mathrm{stat}}$ for each \xJ\ bin is taken to be the
diagonal element of the resulting covariance matrix. The statistical covariance matrices exhibit similar
trends across all \ptone\ and centrality ranges. Nearby \xJ\ bins show
a strong positive correlation that diminishes for bins separated in
\xJ, and is expected from the effects of the procedures for unfolding and mapping to
\xJ. Bins well separated in \xJ\ show an anti-correlation
attributable to the normalisation of \dNdxJ.

\begin{figure}
\centering
\includegraphics[width=0.95\textwidth]{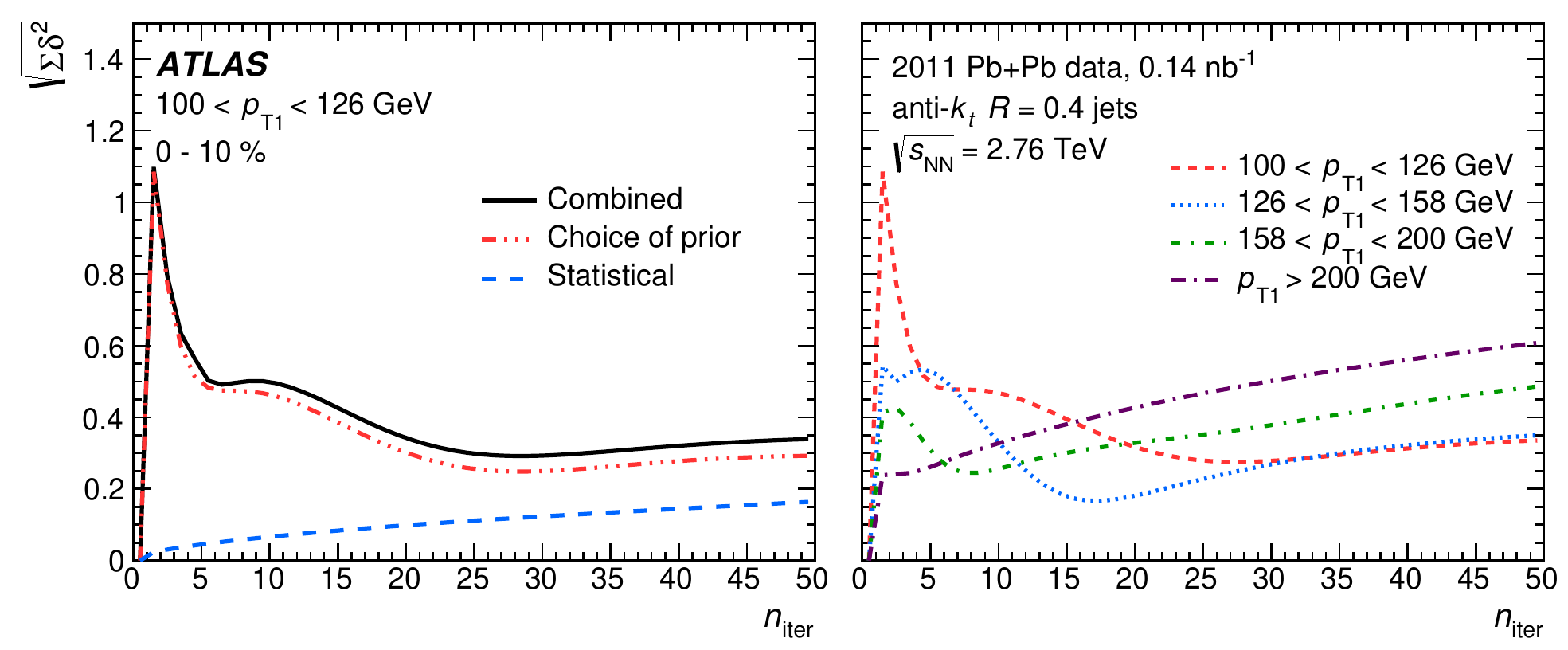}
\caption{Uncertainties sensitive to the number of iterations in the unfolding procedure as a
  function of \niter\ for the 0--10\% centrality interval for $R=0.4$ jets. Left:
  The combination (solid black) of the unfolding (dashed red) and statistical (dotted blue) uncertainty, $\sqrt{\Sigma \delta^2}$ for the $100 < \ptone < 126$~\GeV\ interval. Right: The combined uncertainty for each \ptone\ interval
considered in the measurement.}
\label{fig:niter_choice}
\end{figure}

The left panel of
Figure~\ref{fig:niter_choice} shows $\sqrt{\Sigma \delta^2}$ as a function of \niter\
along with its various contributions for the $100
< \ptone < 126$~\GeV\ range and 0--10\% centrality interval. Since the
unfolding is performed in two dimensions, the value of \niter\ cannot be chosen separately for each range of
\ptone. At higher values of \ptone\ the effects of the unfolding are smaller while the
effects of the statistical fluctuations can be more severe. The right panel of Figure~\ref{fig:niter_choice} shows the total $\sqrt{\Sigma \delta^2}$ for each range of \ptone\ considered in the measurement along with the
total combined over all \ptone\ ranges. The value of
\niter\ for each centrality bin and $R$ value is chosen by considering the
\niter\ dependence of $\sqrt{\Sigma \delta^2}$  for each \ptone\ bin and
selecting a value that maintains comparable uncertainties across all
\ptone\ ranges. The more central bins require the most iterations,
resulting from the larger jet energy resolution in these events. The number of iterations for $R=0.4$ jets is at most 20 for 0--10\% centrality and at the least 6 for 60-80\% centrality. The
$\sqrt{\Sigma \delta^2}$ distributions for $R=0.3$ jets show behaviour similar
 to those for $R=0.4$ jets in the same centrality bin. 

It is possible for a third jet present in the event to be reconstructed as the jet with the 
second highest \pt\ through the experimental resolution. As a check to study the
impact of such effects on the measurement, an alternative response
matrix is constructed where no $\Delta R$ matching is required between
the truth and reconstructed jets. A weighting is applied such that the
\pt\ distribution of the reconstructed third jet matches that observed in the
data. Differences between the unfolded distributions obtained with this
response matrix and the nominal one are observed to be small and well
within the systematic uncertainty associated with the unfolding procedure. 

\begin{figure}
\centering
\includegraphics[width=0.95\textwidth]{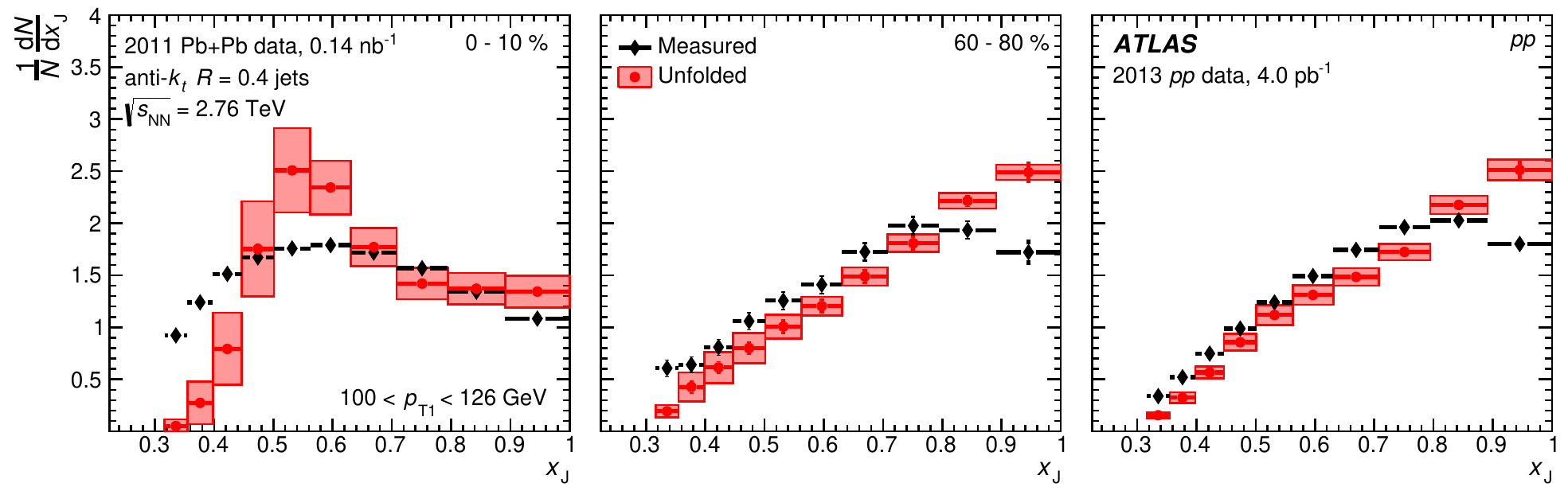}
\caption{The \dNdxJ\ distributions for $R=0.4$ jets before (black) and after (red)
  unfolding for the $100 < \ptone < 126$~\GeV\ interval for the
  \PbPb\ 0--10\% (left) and \PbPb\ 60--80\% (middle) centrality ranges and for \pp\
  collisions (right). Statistical uncertainties are indicated by
  vertical error bars (not visible in most cases). Systematic uncertainties in the unfolded result
  are indicated by the red shaded boxes. }
\label{fig:unfolding_effects}
\end{figure}

The \dNdxJ\ distributions before and after unfolding are shown in
Figure~\ref{fig:unfolding_effects} for central and peripheral \PbPb\
collisions and for \pp\ collisions for jet pairs with $100
< \ptone < 126$~\GeV. The systematic uncertainties indicated contain
all of the contributions to the total systematic uncertainty described
in Section~\ref{sec:systematics}. In the \pp\ and 60--80\% centrality
interval, the resolution effects before unfolding reduce the sharpness of the peak
near $\xJ\sim1$. In the case of the 0--10\% centrality interval, the
effect is to smear out the peak near $\xJ\sim0.5$. The lowest \xJ\ bins
exhibit instability in the unfolding procedure due to the MC sample having too few events 
in this region. However, including this range in the
unfolding improves the stability of the adjacent \xJ\
bins. Thus, after unfolding, only the range $0.32 < \xJ < 1$ is
reported in the results even though pairs with $\pttwo\ >
25$~\GeV\ are included in the measurement. 

\section{Systematic uncertainties}
\label{sec:systematics}

Systematic uncertainties attributed to the response matrix used in the
unfolding arise due to uncertainties in the JES and JER. To account
for these effects, new response matrices are constructed with a
systematically varied relationship between the truth and reconstructed
jet kinematics. The data are then unfolded using the new response and
the result is compared with the nominal.

In the \pp\ data analysis, the JES uncertainty is described by a set
of 11 independent nuisance parameters; these include effects from
uncertainties derived through the \textit{in situ}
calibration~\cite{PERF-2012-01}. In the MC sample used to determine
the calibration, the calorimetric
response to jets initiated by the fragmentation of quarks and gluons
is observed to differ. Potential inaccuracies in the MC sample describing both
this flavour-dependent response and the relative abundances of quark
and gluon jets are accounted for using separate nuisance
parameters. A source of uncertainty related to the
adaptation of the \textit{in situ} calibration derived at $\sqrt{s}=8$~\TeV\ to 2.76~\TeV\
 data is also included. 

In the \PbPb\ data analysis, two additional uncertainties in the JES
are considered. The first accounts for differences between the
detector operating conditions in the \PbPb\ and \pp\
data, which were recorded in 2011 and 2013, respectively. This is derived by using charged-particle tracks reconstructed in the
inner detector to provide an independent check on the JES, which only uses
information from the calorimeter. For each jet, all reconstructed
tracks within $\Delta R < 0.4$ and having $\textit{p}^{\mathrm{trk}}_{\mathrm{T}} >$ 2~\GeV, are matched to the jet and the scalar sum of the track
transverse momenta is evaluated. The ratio of this sum to the jet's \pt\ is
evaluated both in data and in the MC sample, and a double ratio of the two quantities is formed. The double ratio obtained in peripheral \PbPb\ data is compared with that in \pp\ data. The precision of the
comparison is limited by having too few events in the peripheral \PbPb\
data and at high jet \pt, and a \pt- and $\eta$-independent
uncertainty of 1.46\% is assigned to account for potential differences.

The second additional uncertainty is a centrality-dependent JES uncertainty to account for potential
differences in the detector response to quenched jets. This is estimated by comparing the detector response evaluated
in the \textsc{Pythia} and \textsc{Pyquen} MC samples. This estimate
is checked in data using a track-based study similar to the one
described above, but comparing central and peripheral \PbPb\
collisions and accounting for the measured variation of the
fragmentation function with
centrality~\cite{HION-2012-08,CMS-HIN-11-004,CMS-HIN-12-013}. An
uncertainty of  up to 1\% in the most central collisions and decreasing linearly 
with centrality percentile to 0\% in the 60--80\% centrality class is assigned.

The uncertainty attributed to the JER is obtained by adding Gaussian fluctuations to each reconstructed jet \pt\ value
when populating the response matrix. The magnitude of this uncertainty
is fixed by a comparison of the data and MC descriptions of the JER
in 8~\TeV\ data~\cite{PERF-2011-04}. Since the MC
sample is constructed using the data overlay procedure, it is
expected that the centrality dependence of the JER should be well
described in the MC sample. This is checked by studying the distribution of UE fluctuations using random,
jet-sized groups of calorimeter towers in  \PbPb\ data. The standard deviations of
these distributions describe the typical UE contribution beneath a
jet. The centrality dependence of the UE fluctuations is compared to
that of the JER in the MC sample, and a systematic uncertainty is included to account
for the observed differences. As expected, these differences are much smaller than
the centrality-independent contribution to the JER uncertainty.

The data-driven estimates of the JES and JER uncertainties described
above are derived using $R=0.4$ jets. Additional uncertainties
are included in the $R=0.3$ jet measurement to account for potential
differences between data and the MC sample in the relative energy
scale of $R=0.3$ jets with respect to $R=0.4$ jets. These uncertainties
are estimated from a study that matched jets reconstructed with the
two $R$ values and compared the means of the
$p_{\mathrm{T}}^{R=0.3} / p_{\mathrm{T}}^{R=0.4}$ distributions in
data and the MC sample. Differences may arise between the data and MC
sample from differences in the calorimetric response or because the
jets in the two samples have different internal structure. The
contribution of the latter is constrained by using existing
jet shape measurements~\cite{STDM-2010-10}. An uncertainty in the energy scale is
applied to account for residual differences, which are 1.5\%
at the lowest \pt\ and decrease sharply as a function of \pt\ to a
limiting value of 0.3\% at high \pt. A similar study comparing the
variances of the $p_{\mathrm{T}}^{R=0.3} / p_{\mathrm{T}}^{R=0.4}$
distributions is used to constrain the uncertainty in the relative
resolution. This uncertainty is applied in the $R=0.3$ jet
measurement in the same fashion as the other JER uncertainties
described above. Although larger than the centrality-dependent
contribution, it is also much smaller than the centrality-independent contribution.

As the response matrix is sparsely populated
(containing $40^4$ bins), statistical fluctuations could
introduce instabilities in the unfolding. To evaluate the sensitivity
to such effects, along with any other defects in the response, a new
response matrix is constructed as a factorised product of single-jet
response distributions, i.e. assuming the responses in \ptone\ and
\pttwo\ are independent. The data are unfolded using this new
response and the differences between the unfolded distributions are taken
as a systematic uncertainty. Systematic uncertainties in the unfolding due to the choice of prior are estimated as described in the previous section and are also included.

Uncertainties due to the correction for the combinatoric effects
described in Section~\ref{sec:analysis} affect the number of jet pairs before the unfolding and are thus included
as additional contributions to the previously described statistical uncertainties in the data. These include statistical uncertainties in
$\varepsilon$ and the uncertainties in the values of the fit
parameters $c_3$ and $c_4$, accounting for their
covariance. Uncertainties in the normalisation are
estimated by varying the region of $\Delta \phi$ used to estimate $Y$
from 1.0--1.4 to 1.1--1.5.  The uncertainty due to this
correction is smaller than the other uncertainties in all \pt\ and centrality
bins, and is only greater than 5\% at values of $\xJ < 0.4$. This correction was not applied to the \pp\ data so there is no corresponding systematic uncertainty.  

\begin{figure} \centering
\includegraphics[width=0.95\textwidth]{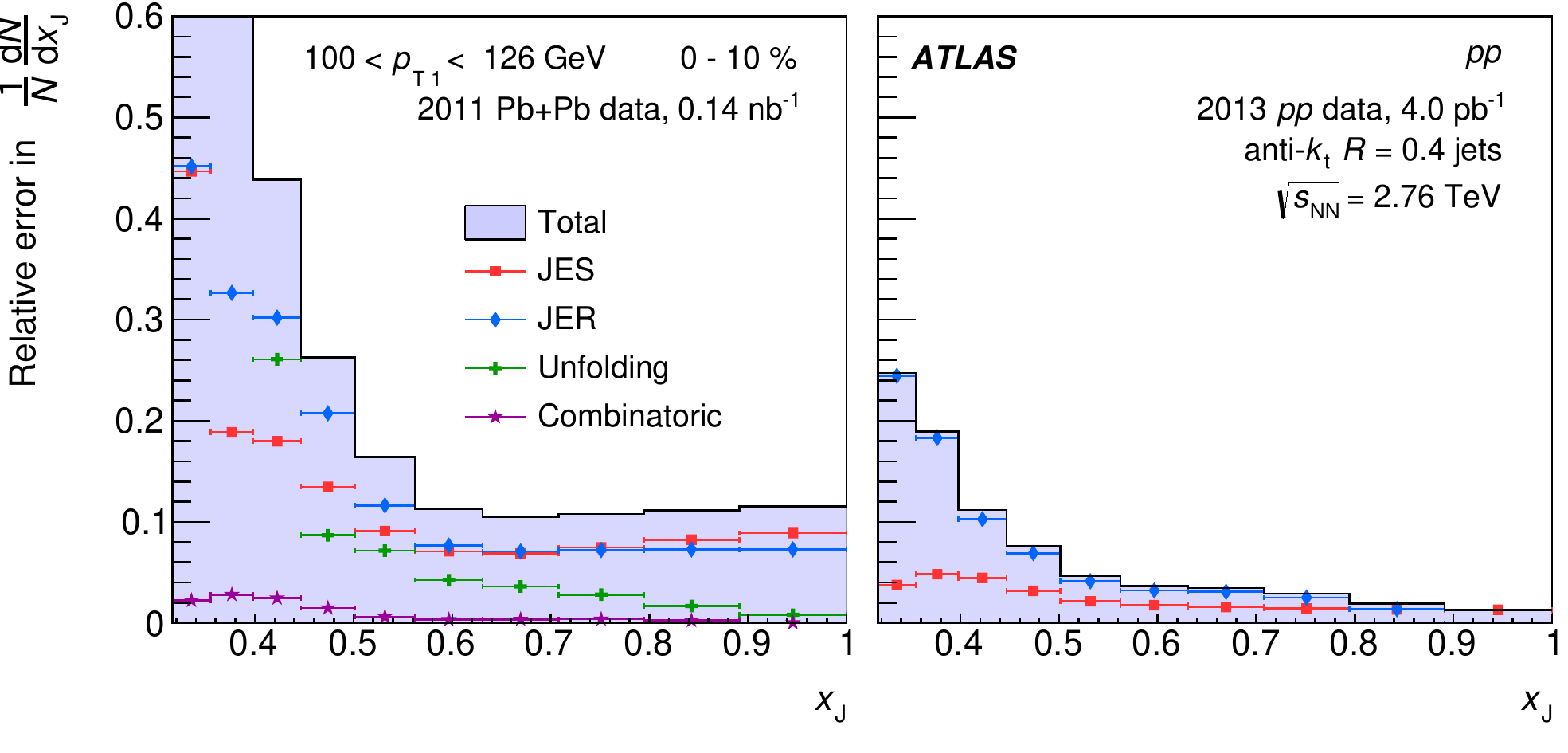}
 \caption{The total systematic uncertainty and its various components
for $100 < \ptone < 126$~\GeV\ for $R=0.4$ jets in \PbPb\ collisions with 0--10\%
centrality (left) and \pp\ collisions (right). In the figure on the left the first two bins are off scale with bins centers of \xJ=0.34 and 0.38 and bins contents of 1.25 and 0.75, respectively.}
\label{fig:total_uncertainty_pt0}
\end{figure}

The breakdown of different contributions to the total systematic
uncertainty is shown in the $100 < \ptone <  126$~\GeV\ range for the
0--10\% centrality interval and for \pp\ collisions in
Figure~\ref{fig:total_uncertainty_pt0}. Each contribution to the uncertainty, and thus the total uncertainty,
tends to decrease with increasing \xJ. The total uncertainty at $\xJ\sim1$ reaches approximately 12\% in
most \ptone\ and centrality bins in the \PbPb\ data. For $\xJ < 0.4$, the relative
uncertainty becomes large, but this region represents only a small
contribution to the total \dNdxJ\ distribution. The JER uncertainty is
the largest contribution. In the \PbPb\ data it reaches values of
approximately 10\% and 15\% at $\xJ\sim1$ and $\xJ=0.5$,
respectively. The JES contributions are the second largest
contribution to the uncertainties, typically
between 5\% and 10\%. In the most central bins the
unfolding uncertainty can become as large as the JES
contribution. The contributions to the uncertainty in the other
centrality intervals and in the \pp\ data
follow trends similar to those described for the 0--10\% centrality interval, but the
magnitudes are smaller in more peripheral collisions. In the \pp\ data
they are typically smaller by a factor of two compared to the 0--10\% \PbPb\ data. The uncertainties for the
$R=0.3$ result follow the same trends as those for the $R=0.4$
result but are slightly larger due to the two additional sources
included in that measurement to describe the relative energy scale and
resolution between the two $R$ values.

\section{Results}
\label{sec:results}
\begin{figure}
\centering
\includegraphics[width=0.9\textwidth]{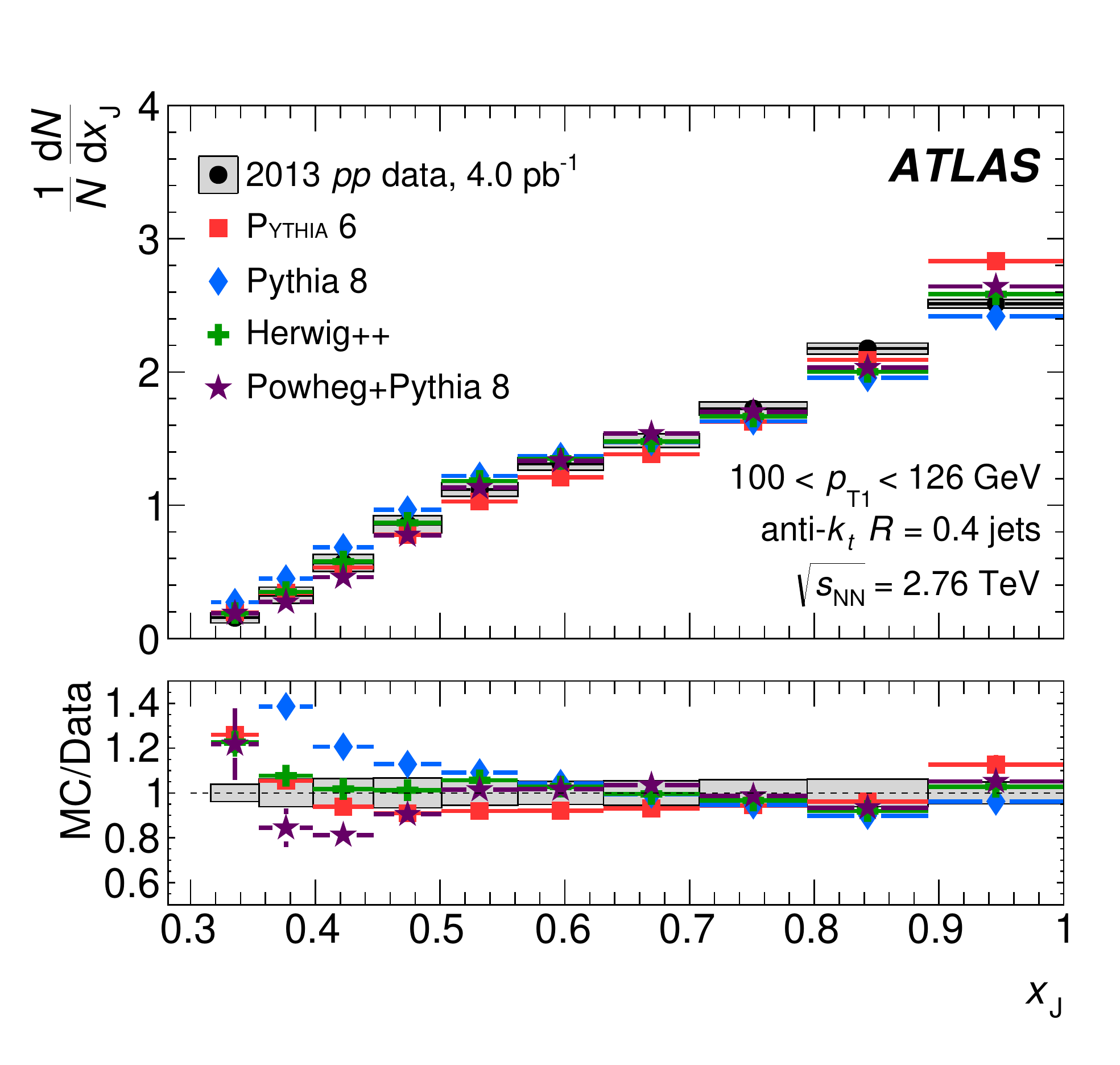}
\caption{The \dNdxJ\ distribution for $R=0.4$ jets in \pp\ collisions for the $100 < \ptone <
  126$~\GeV\ interval is shown in black points with the grey shaded boxes
  indicating the systematic uncertainties. Also shown are results
  obtained from various MC event generators: \pythiaSix\ (red squares),
  \pythiaEight\ (blue diamonds), \herwigpp\ (green crosses)
  and Powheg+\pythiaEight\ (purple stars). The ratio of each MC
  result to the data is shown in the bottom panel where the systematic
uncertainties of the data are indicated by a shaded band centred at unity.}
\label{fig:compare_MC}
\end{figure}

The unfolded \dNdxJ\ distribution in \pp\ collisions for $100 < \ptone <
  126$~\GeV\ is shown in Figure~\ref{fig:compare_MC}. Also shown are the corresponding
  distributions obtained from the \pythiaSix\ sample used in the
  MC studies and also from \pythiaEight\  using the AU2 tune and \herwigpp\ \cite{Bahr:2008pv} with the
  UE-EE-3 \cite{Gieseke:2012ft} tune. An additional sample, referred to as
  Powheg+\pythiaEight\, is generated using Powheg-Box
  2.0~\cite{Nason:2004rx,Frixione:2007vw,Alioli:2010xd}, which is
  accurate to next-to-leading order in perturbative QCD, and
  interfaced with \pythiaEight\ to provide a description of the parton
  shower and hadronisation. All samples used the CTEQ6L1 PDF set
  \cite{Pumplin:2002vw} except the Powheg+\pythiaEight, which used the
  CT10 PDF set \cite{Lai:2010vv}. All four models describe the data
  fairly well with the \herwigpp\ and Powheg+\pythiaEight\ showing the
  best agreement over the full \xJ\ range.

\begin{figure}
\centering
\includegraphics[width=0.9\textwidth]{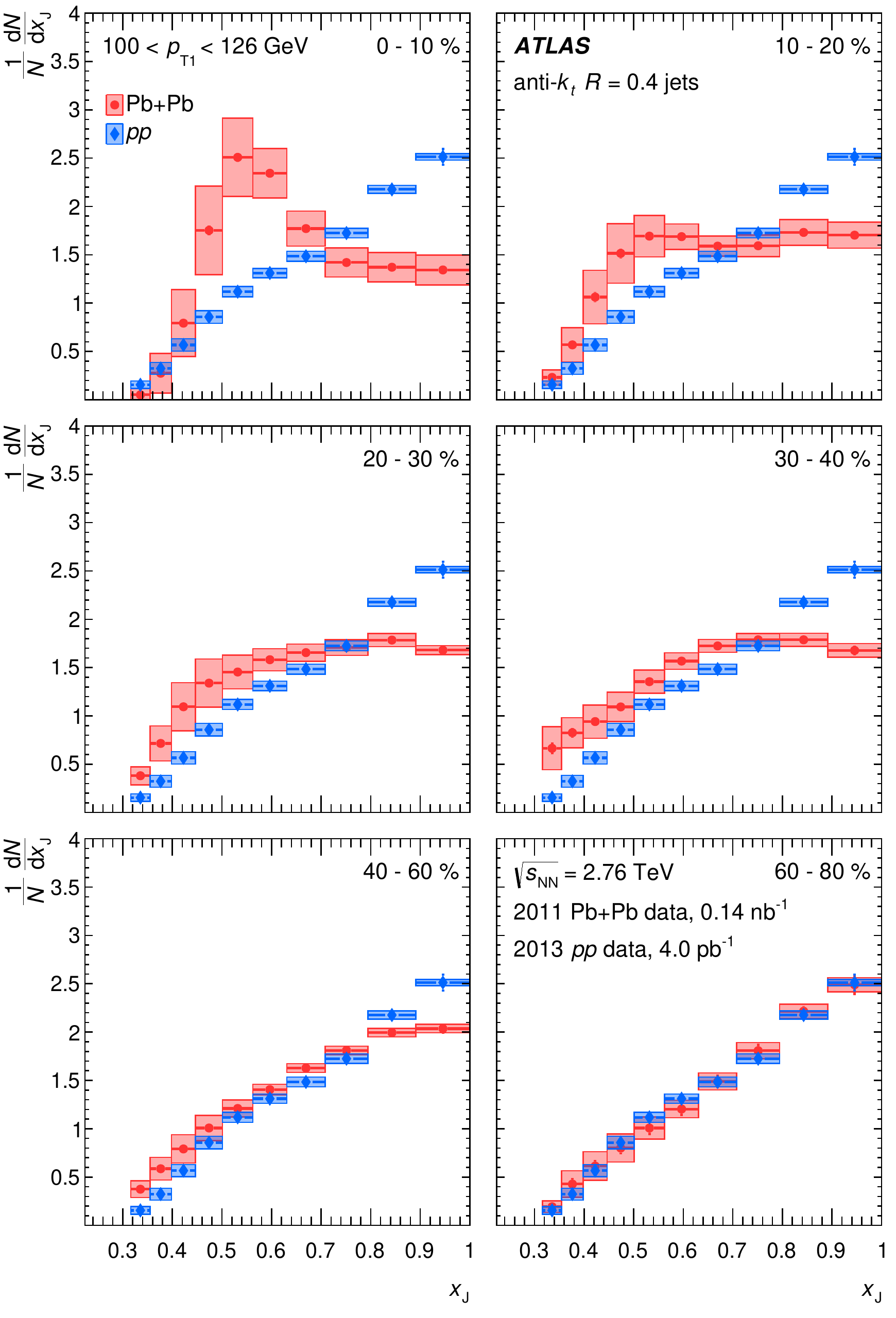}
\caption{The \dNdxJ\ distributions for jet pairs with $100 < \ptone <
  126$~\GeV\ for different collision centralities for $R=0.4$ jets. The \PbPb\ data are
  shown in red circles, while the \pp\ distribution is shown for comparison in
blue diamonds, and is the same in all panels. Statistical uncertainties are
indicated by the error bars while systematic uncertainties are shown
with shaded boxes.}
\label{fig:xj_cent_fixed_pt}
\end{figure}
The unfolded \dNdxJ\ distributions in \PbPb\ collisions are shown in Figure~\ref{fig:xj_cent_fixed_pt}, for jet pairs with  $100 < \ptone <
  126$~\GeV\ for different centrality intervals. The distribution in
  \pp\ collisions is shown on each panel for comparison. In the
  60--80\% centrality bin, where the effects of quenching are expected
  to be the smallest, the \PbPb\ data are consistent with
  the \pp\ data. In more central Pb+Pb collisions, the distributions become
  significantly broader than that in \pp\ collisions and the peak at $\xJ\sim1$, 
  corresponding to nearly symmetric dijet events, is reduced. At lower centrality percentiles the distribution becomes almost constant over the range $0.6 \lesssim \xJ
  \lesssim 1$, and develops a peak at $\xJ\sim0.5$ in the 0--10\% centrality interval.

\begin{figure}
\centering
\includegraphics[width=0.9\textwidth]{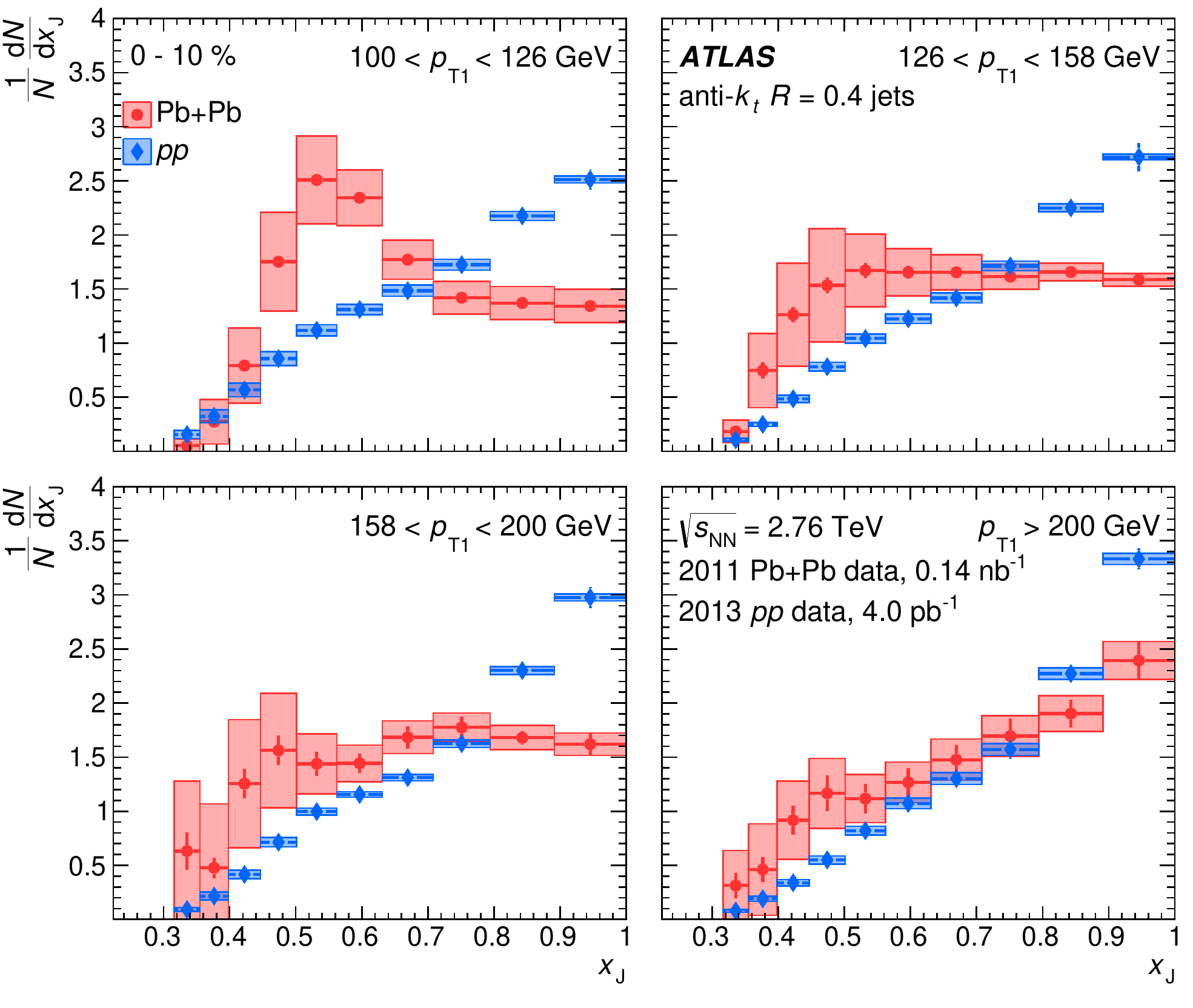}
\caption{The \dNdxJ\ distributions for $R=0.4$ jets with different selections on \ptone,
  shown for the 0--10\% centrality bin (red circles) and for \pp\ (blue diamonds). Statistical uncertainties are
indicated by the error bars while systematic uncertainties are shown
with shaded boxes.}
\label{fig:xj_pt_fixed_cent}
\end{figure}

Figure~\ref{fig:xj_pt_fixed_cent} shows the \dNdxJ\ distributions for 0--10\% centrality \PbPb\
collisions and \pp\ collisions for different selections on \ptone. In
\pp\ collisions, the \xJ\ distribution becomes increasingly narrow with increasing \ptone,
indicating that higher-\pt\ dijets tend to be better balanced in
momentum (fractionally). At higher \ptone, the \xJ\ distribution begins to fall more
steeply from $\xJ\sim1$, but appears to flatten at intermediate values
of $\xJ$. The modifications observed in the \PbPb\ data lessen with
increasing \ptone\ and for jet pairs with $\ptone > 200$~\GeV\ the maximum
at $\xJ\sim1$ is restored.

\begin{figure}
\centering
\includegraphics[width=0.9\textwidth]{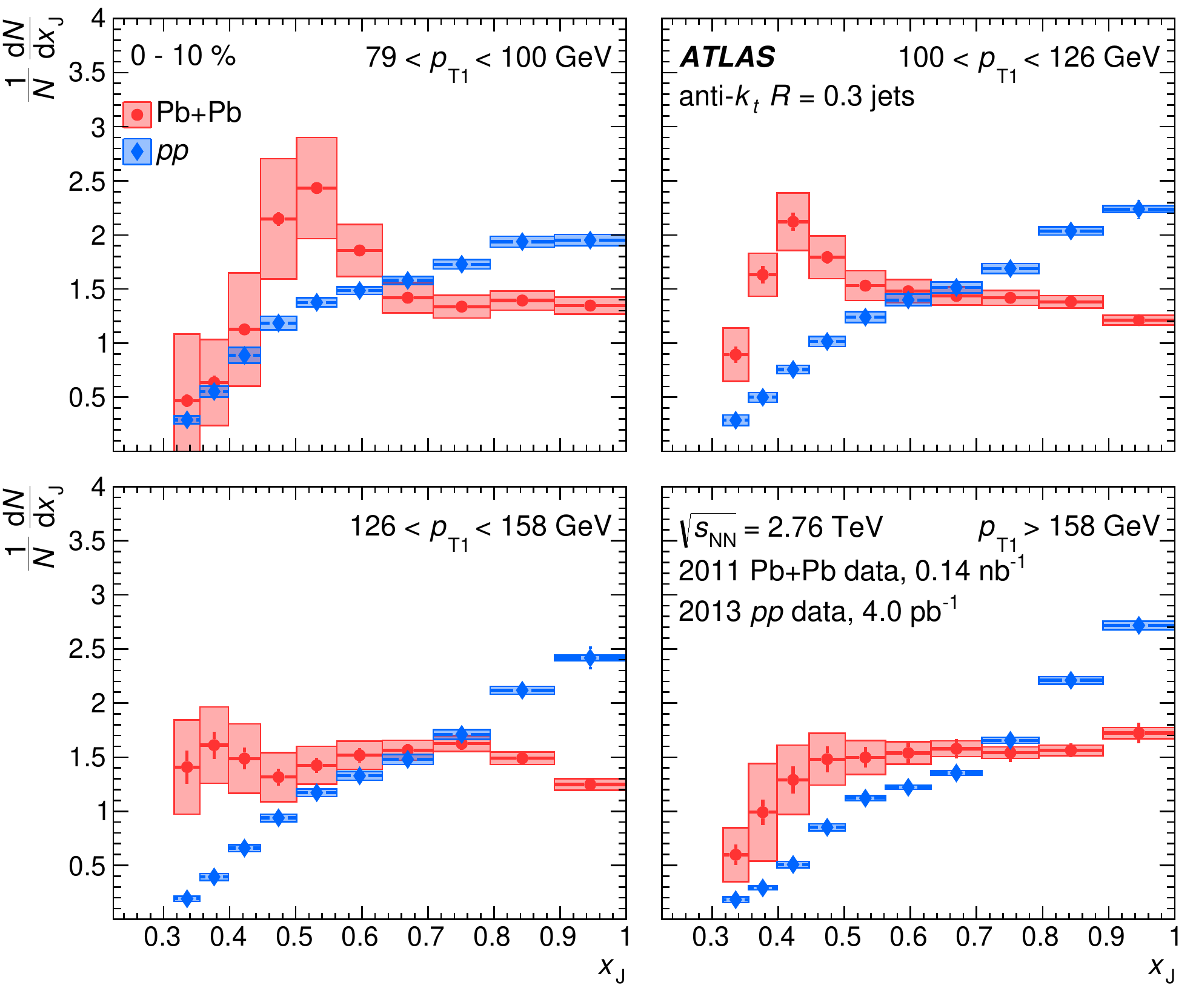}
\caption{The \dNdxJ\ distributions for $R=0.3$ jets with different selections on \ptone,
  shown for the 0--10\% centrality bin (red circles) and for \pp\ (blue diamonds). Statistical uncertainties are
indicated by the error bars while systematic uncertainties are shown
with shaded boxes.}
\label{fig:xj_cent_fixed_pt_03}
\end{figure}
\begin{figure}
\centering
\includegraphics[width=0.9\textwidth]{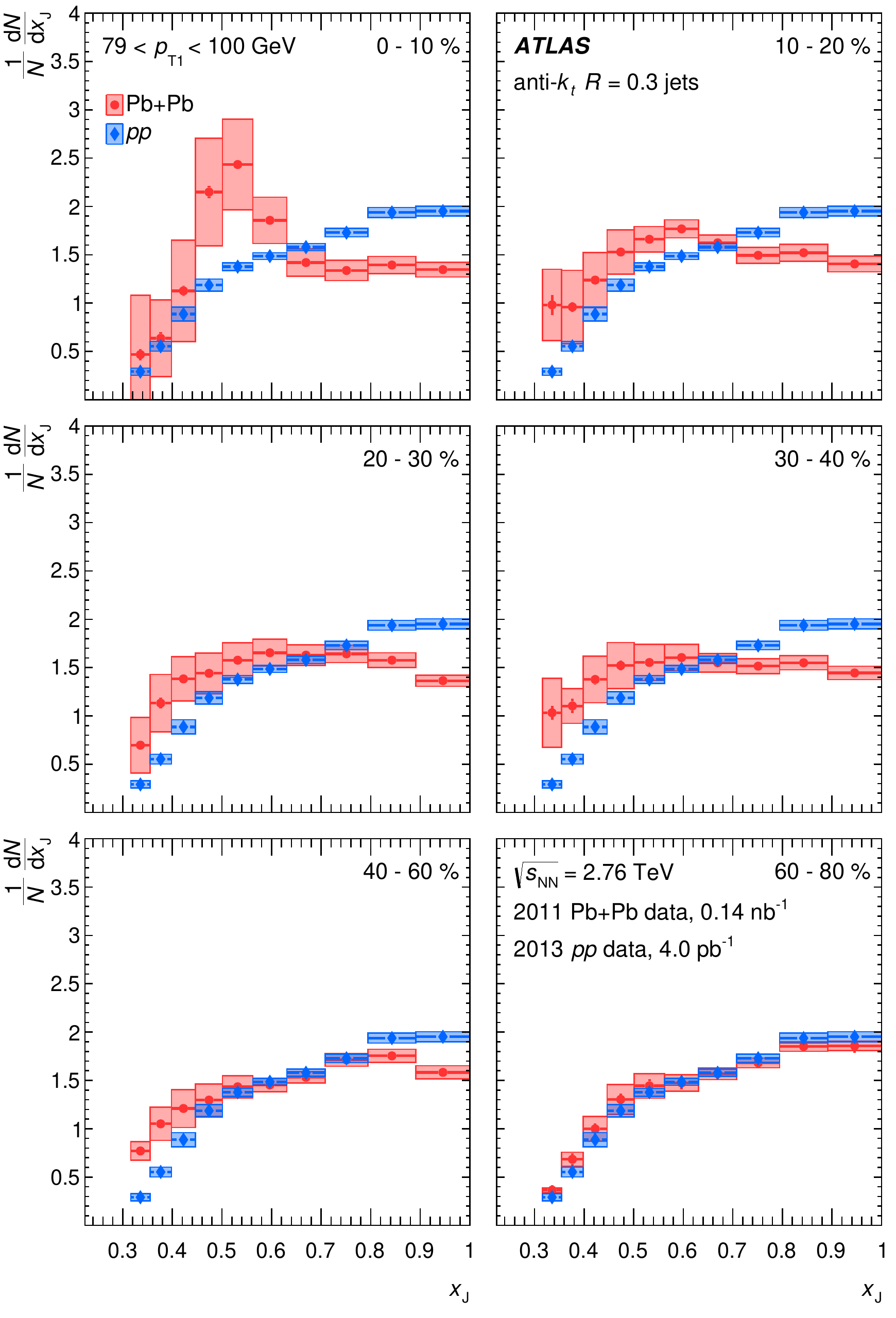}
\caption{The \dNdxJ\ distributions for jet pairs with $79 < \ptone <
  100$~\GeV\ for different collision centralities for $R=0.3$ jets. The \PbPb\ data are
  shown in red circles, while the \pp\ distribution is shown for comparison in
blue diamonds, and is the same in all panels. Statistical uncertainties are
indicated by the error bars while systematic uncertainties are shown
with shaded boxes.}
\label{fig:xj_pt_fixed_cent_03}
\end{figure}

The distributions for $R=0.3$ jets are also shown for the
0--10\% centrality interval and for \pp\ collisions for different
\ptone\ ranges in Figure~\ref{fig:xj_cent_fixed_pt_03}. The \pt\ of an $R=0.3$ jet is
generally lower than that of an $R=0.4$ jet originating from the same
hard scattering, and thus features observed in the \dNdxJ\ distributions
for $R=0.4$ jets are expected to appear at lower values of \ptone\
for $R=0.3$ jets. To facilitate a comparison between results obtained
with the two $R$ values, the $R=0.3$ jet results include an additional
\ptone\ interval, $79 < \ptone < 100$~\GeV. The differences between the \PbPb\ and
\pp\ \dNdxJ\ distributions are qualitatively similar to those observed
for $R=0.4$ jets. Figure \ref{fig:xj_pt_fixed_cent_03} shows the \dNdxJ\ distributions for  $79 < \ptone <
  100$~\GeV\ for different collision centralities but for jets
reconstructed with $R=0.3$. This indicates that the trends present in \ptone\
and centrality are
robust with respect to the UE and that UE effects are properly
accounted for by the combinatoric subtraction and unfolding procedures
applied in the data analysis. The distributions are
flatter for $R=0.3$ jets in all \pt\ and centrality bins, including in
\pp\ collisions. This is consistent with the expectation that the
\ptonepttwo\ correlation is weaker for smaller-$R$ jets due to the
effects of parton radiation outside the nominal jet cone.

\FloatBarrier
\section{Conclusion}
\label{sec:conclusion}
This \papertype\ presents a measurement of dijet \xJ\ distributions in 4.0\invpb\ of \pp\ and 0.14\invnb\ of \PbPb\ collisions at
$\sqrtsnn=2.76$~\TeV. The measurement is performed differentially in
leading-jet transverse momentum, \ptone, and in collision
centrality using data from the ATLAS detector at the LHC. The measured distributions are unfolded to account for
the effects of experimental resolution and inefficiencies on the two-dimensional
\ptonepttwo\ distributions and then projected into bins of fixed ratio
$\xJ=\pttwo/\ptone$. The distributions show a larger contribution of asymmetric dijets in \PbPb\ data compared to that in \pp\ data, a feature that
becomes more pronounced in more central collisions and is consistent with expectations of medium-induced energy loss
due to jet quenching. In the 0--10\% centrality bin for $100 < \ptone <
126$~\GeV, the \xJ\ distribution develops a significant peak at
$\xJ\sim0.5$ indicating that the most probable configuration for
dijets is for them to be highly unbalanced. This is in sharp contrast
to the situation in the \pp\ data where the most probable values are
near $\xJ\sim1$. The centrality-dependent
modifications evolve smoothly from central to peripheral collisions,
and the results in the 60--80\% centrality bin and the \pp\ data are consistent. At larger values of \ptone\, the \xJ\ distributions
are observed to narrow and the differences between the distributions
in central \PbPb\ and \pp\ collisions lessen. This is qualitatively consistent with
a picture in which the fractional energy loss decreases with increasing jet \pt.
The features in the data are compatible with those
observed in previous measurements of dijets in \PbPb\ collisions by the ATLAS and CMS collaborations, 
however, the trends in this measurement are more prominent due to the application of
the unfolding procedure. This result constitutes an important
benchmark for theoretical models of jet quenching and the dynamics of
relativistic heavy-ion collisions.
\section*{Acknowledgements}

We thank CERN for the very successful operation of the LHC, as well as the
support staff from our institutions without whom ATLAS could not be
operated efficiently.

We acknowledge the support of ANPCyT, Argentina; YerPhI, Armenia; ARC, Australia; BMWFW and FWF, Austria; ANAS, Azerbaijan; SSTC, Belarus; CNPq and FAPESP, Brazil; NSERC, NRC and CFI, Canada; CERN; CONICYT, Chile; CAS, MOST and NSFC, China; COLCIENCIAS, Colombia; MSMT CR, MPO CR and VSC CR, Czech Republic; DNRF and DNSRC, Denmark; IN2P3-CNRS, CEA-DSM/IRFU, France; SRNSF, Georgia; BMBF, HGF, and MPG, Germany; GSRT, Greece; RGC, Hong Kong SAR, China; ISF, I-CORE and Benoziyo Center, Israel; INFN, Italy; MEXT and JSPS, Japan; CNRST, Morocco; NWO, Netherlands; RCN, Norway; MNiSW and NCN, Poland; FCT, Portugal; MNE/IFA, Romania; MES of Russia and NRC KI, Russian Federation; JINR; MESTD, Serbia; MSSR, Slovakia; ARRS and MIZ\v{S}, Slovenia; DST/NRF, South Africa; MINECO, Spain; SRC and Wallenberg Foundation, Sweden; SERI, SNSF and Cantons of Bern and Geneva, Switzerland; MOST, Taiwan; TAEK, Turkey; STFC, United Kingdom; DOE and NSF, United States of America. In addition, individual groups and members have received support from BCKDF, the Canada Council, CANARIE, CRC, Compute Canada, FQRNT, and the Ontario Innovation Trust, Canada; EPLANET, ERC, ERDF, FP7, Horizon 2020 and Marie Sk{\l}odowska-Curie Actions, European Union; Investissements d'Avenir Labex and Idex, ANR, R{\'e}gion Auvergne and Fondation Partager le Savoir, France; DFG and AvH Foundation, Germany; Herakleitos, Thales and Aristeia programmes co-financed by EU-ESF and the Greek NSRF; BSF, GIF and Minerva, Israel; BRF, Norway; CERCA Programme Generalitat de Catalunya, Generalitat Valenciana, Spain; the Royal Society and Leverhulme Trust, United Kingdom.

The crucial computing support from all WLCG partners is acknowledged gratefully, in particular from CERN, the ATLAS Tier-1 facilities at TRIUMF (Canada), NDGF (Denmark, Norway, Sweden), CC-IN2P3 (France), KIT/GridKA (Germany), INFN-CNAF (Italy), NL-T1 (Netherlands), PIC (Spain), ASGC (Taiwan), RAL (UK) and BNL (USA), the Tier-2 facilities worldwide and large non-WLCG resource providers. Major contributors of computing resources are listed in Ref.~\cite{ATL-GEN-PUB-2016-002}.

\printbibliography
\newpage
\begin{flushleft}
{\Large The ATLAS Collaboration}

\bigskip

M.~Aaboud$^\textrm{\scriptsize 137d}$,
G.~Aad$^\textrm{\scriptsize 88}$,
B.~Abbott$^\textrm{\scriptsize 115}$,
J.~Abdallah$^\textrm{\scriptsize 8}$,
O.~Abdinov$^\textrm{\scriptsize 12}$,
B.~Abeloos$^\textrm{\scriptsize 119}$,
S.H.~Abidi$^\textrm{\scriptsize 161}$,
O.S.~AbouZeid$^\textrm{\scriptsize 139}$,
N.L.~Abraham$^\textrm{\scriptsize 151}$,
H.~Abramowicz$^\textrm{\scriptsize 155}$
% ...
\bigskip
\\
% ...
$^{8}$ Department of Physics, The University of Texas at Arlington, Arlington TX, United States of America\\
% ...
$^{12}$ Institute of Physics, Azerbaijan Academy of Sciences, Baku, Azerbaijan\\
% ...
$^{88}$ CPPM, Aix-Marseille Universit{\'e} and CNRS/IN2P3, Marseille, France\\
% ...
$^{115}$ Homer L. Dodge Department of Physics and Astronomy, University of Oklahoma, Norman OK, United States of America\\
% ...
$^{119}$ LAL, Univ. Paris-Sud, CNRS/IN2P3, Universit{\'e} Paris-Saclay, Orsay, France\\
% ...
$^{137}$ $^{(a)}$ Facult{\'e} des Sciences Ain Chock, R{\'e}seau Universitaire de Physique des Hautes Energies - Universit{\'e} Hassan II, Casablanca; $^{(b)}$ Centre National de l'Energie des Sciences Techniques Nucleaires, Rabat; $^{(c)}$ Facult{\'e} des Sciences Semlalia, Universit{\'e} Cadi Ayyad, LPHEA-Marrakech; $^{(d)}$ Facult{\'e} des Sciences, Universit{\'e} Mohamed Premier and LPTPM, Oujda; $^{(e)}$ Facult{\'e} des sciences, Universit{\'e} Mohammed V, Rabat, Morocco\\
% ...
$^{139}$ Santa Cruz Institute for Particle Physics, University of California Santa Cruz, Santa Cruz CA, United States of America\\
% ...
$^{151}$ Department of Physics and Astronomy, University of Sussex, Brighton, United Kingdom\\
% ...
$^{155}$ Raymond and Beverly Sackler School of Physics and Astronomy, Tel Aviv University, Tel Aviv, Israel\\
% ...
$^{161}$ Department of Physics, University of Toronto, Toronto ON, Canada\\
% ...
$^{a}$ Also at Department of Physics, King's College London, London, United Kingdom\\
$^{b}$ Also at Institute of Physics, Azerbaijan Academy of Sciences, Baku, Azerbaijan\\
$^{c}$ Also at Novosibirsk State University, Novosibirsk, Russia\\
$^{d}$ Also at TRIUMF, Vancouver BC, Canada\\
$^{e}$ Also at Department of Physics {\&} Astronomy, University of Louisville, Louisville, KY, United States of America\\
% ...
\end{flushleft}


\end{document}